\documentstyle [12pt]{article}
\advance\hoffset by -.5 in
\advance\voffset by -1 in 
\begin{document}

\def\pz{\p_z}
\def\a{\alpha}
\def\b{\beta}
\def\g{\gamma}
\def\d{\delta}
\def\e{\epsilon}
\def\x{\xi}            
\def\f{\phi}
\def\j{\psi}
\def\t{\tau}
\def\L{\Lambda}
\def\l{\lambda}
\def\p{\partial}
\def\o{\omega}
\def\O{\Omega}
\def\z{\zeta}
\def\Tau{{\rm T}}
\def\cA{{\cal A}}
\def\tcA{\tilde{\cal A}}
\def\tW{{\bf\tilde W}}
\def\cK{{\cal K}}
\def\tf{\tilde{f}}
\def\tg{\tilde{g}}
\def\th{\tilde{h}}
\def\bL{{\bf L}}
\def\bM{{\bf M}}
\def\bN{{\bf N}}
\def\bP{{\bf P}}
\def\bW{{\bf W}}
\def\rarr{\rightarrow}
\def\mp{\mapsto}
\def\hL{\hat{L}}
\def\hw{\hat w}
\def\res{{\rm res}}
\def\tr{{\rm tr}~}
\def\tres{\tr\res}
\def\cF{{\cal F}}
\def\et{\eta}
\def\cV{{\cal V}}
\def\s{\sum}
\def\cL{{\cal L}}
\def\cH{{\cal H}}
\def\hw{\hat{w}}
\def\bg{{\bf g}}
\def\un{\underline}
\begin{center}
\begin{bf}
\begin{Large}
Modified KP and Discrete KP \\
\end{Large}
\vspace{.5in}
L.A.Dickey\\
\end{bf}
\vspace{.3in}
University of Oklahoma, Norman, OK 73019 \\
e-mail: ldickey@math.ou.edu \\
\vspace{.3in}
February 1999
\end{center}

\vspace{.5in}
The aim of this note is to show that the so-called discrete KP, or 
1-Toda lattice
hierarchy (see Adler and van Moerbeke [AvM 98], Ueno and Takasaki (UT 84) )
is the same as a properly defined modified KP hierarchy. Virtually, the
relationship between them is of the same sort as between the GD (or the
$n$th KdV) hierarchy based on $n$th order differential operators and the
hierarchy based on matrix first-order operators. This relationship is
given by the Drinfeld-Sokolov reduction [DS 81]. In [D 81] (see also [D 91] 
and [D 97]) the following version of this reduction was exploited. Linear 
transformations in the space of differential and pseudodifferential operators 
(for example, the Adler mapping) can be represented by matrices in a basis of 
$\p^k$. Thus, the matrices in the DS reduction must be treated as linear 
transformations not in an abstract linear space but in a concrete space of 
pseudodifferential operators, which gives nice formulas.
Similarly, here we deal with the space of series in $z$ and $z^{-1}$ where
the Baker functions live, or, alternatively, the space of series in $(\p,\p^{-1})$
for dressing operators. Linear transformations of this space involved 
in the modified KP can be represented by matrices of infinite order, and then 
we get the discrete KP. \\

There exists several definitions of the modified KP. All of them are trying to
imitate the relationship between the KdV (GD) and the modified KdV
hierarchy. A complete analogy is impossible since a pseudodifferential
operator cannot be factorized in a product of infinitely many first order 
linear differential operators, a kind of the Miura transform.
Palliative solutions of this problem in the existing definitions have,
in our opinion, serious shortcomings both, aesthetical and practical.
First of all, they are lacking symmetry. The factors are of different
character, some of them are first order differential and some pseudodifferential
operators, or all pseudodifferential, but in this latter case there are introduced
manifestly too many variables. Some of definitions contain an artificial 
additional integer parameter. 

{\em We believe that the discrete KP (1-Toda lattice) hierarchy is the best
generalization of the modified KdV and we try to show this in the
present article. This is its main point.}\\ 

The modified KP is obtained by a successive extension of a simple, ordinary KP
joining to it new and new unknown functions (``fields'') 
by B\"acklund (or Darboux) transformations. 
Independent variables (``times'') remain the same. The old fields, which already 
existed before an act of extension, just do not feel this act. This is a 
difference between this type of extension and that in the case of the 
Zakharov-Shabat multipole hierarchy (see [D 94]) where an additional 
pole increased both, the number of fields and the number of times, so
that old fields depended on new times, too.

The paper is self-contained, all definitions are given.\\

{\bf 1. Modified KdV.}\\ 

The modified $n$th KdV (or GD) hierarchy is
well-known. We will follow our paper [D 93] in its description.

Let $v_1,...,v_n$ be generators of a differential algebra ${\cal A}_v$
with a relation $v_1+...+v_n=0$. We shall also use $v_k$ with any integer 
subscript $k$ assuming that $v_{k+n}=v_k$. Let
$$L_i:=(\partial +v_{i-1})...(\partial +v_1)(\partial +v_n)...(\partial +v_{i})
=(\partial +v_{i+n-1})...(\partial +v_{i})\eqno{(1.1)}$$

Each $L_i$ can also be represented as
$$ L_i=\partial^n+u_1^{[i]}\partial^{n-1}+...+u_n^{[i]}.$$
The coefficients $\{u_k^{[i]}\}$, $k=1,...,n$ are elements of ${\cal A}_v$
$$ \hspace{.2in} u_k^{[i]}=F_k(v_i,...,v_{i+n-1}),\:k=1,...,n.$$
These functions define the Miura transformation. They determine an embedding
of the differential algebra ${\cal A}_{u^{[i]}}$ of all differential
polynomials in {$u_k^{[i]}$} into ${\cal A}_v$,
$${\cal A}_{u^{[i]}}\subset {\cal A}_v.$$ It is obvious that $$(\partial +
v_i)L_i=L_{i+1}(\partial +v_i)\eqno{(1.2)}$$ whence
$$(\partial +v_i)L_i^{k/n}=L_{i+1}^{k/n}(\partial +v_i)$$ and
$$ (\partial +v_i)(L_i^{k/n})_+-(L_{i+1}^{k/n})_+(\partial +v_i)=
-(\partial +v_i)(L_i^{k/n})_-+(L_{i+1}^{k/n})_-(\partial +v_i).$$ As usual, 
the subscript $+$
denotes the positive (differential) part of a pseudodifferential
operator, and the subscript $-$ the rest of it. The r.-h.s. is a at most 
zero-order pseusodifferential operator ($\Psi$DO) while the l.-h.s. is a 
differential operator. Hence, we proved:\\

{\bf Lemma 1.1.} {\sl
$$ (\partial +v_i)(L_i^{k/n})_+-(L_{i+1}^{k/n})_+(\partial +v_i)$$ is a
zero-order differential operator, i.e., a function (an element of ${\cal
A}_v$).}\\

{\bf Corollary 1.2.} {\sl The system of equations for $v_k$:
$$\p_kv_i=(L_{i+1}^{k/n})_+(\partial +v_i)-(\partial +v_i)(L_i^{k/n})_+,~~
\p_k=\p/\p t_k,\eqno{(1.3)}$$ where $t_k$ are parameters, and $L_i$
are defined by (1.1), makes sense. }\\

{\bf Definition 1.3.} {\sl  The system (1.3), (1.1) is called the $n$th
modified KdV, mKdV (or mGD).}\\

{\bf Proposition 1.4.}  {\sl The system (1.3), (1.1) guarantees that all
$L_i$ satisfy the n-th KdV equation.}\\

{\em Proof.} We have 
$$\partial_kL_i=\partial_k(\partial +v_{i+n-1})...(\partial +v_{i})$$ $$=\sum_{l=i}
^{i+n-1}(\partial +v_{i+n-1})...(\partial +v_{l+1})
\left((L_{l+1}^{k/n})_+(\partial +v_l)-(\partial+v_l)(L_{l}^{k/n})_+\right) (\p+v_{l-1})...
(\partial +v_{i})$$ $$=(L_i^{k/n})_+L_i-L_i(L_i^{k/n})_+=[(L_i^{k/n})_+,L_i].
~\Box$$

Each operator $L_i$ can be represented in a dressing form $L_i=\hat w_i\p^n
\hat w_i^{-1}$ where $\hat{w}_i=\sum_0^{\infty}w_{i,l}\partial^{-l}$ with
$w_{i,0}=1$. Operators $\hw_i$ are determined up to multiplication on the right
by constant series $c_i(z)=\sum_0^{\infty}c_{i,l}\partial^{-l}$.\\

{\bf Lemma 1.5.} {\em Properly choosing $c_i(z)$, one can always achieve the
equality}: $$(\p+v_i)\cdot \hw_{i}=\hw_{i+1}\cdot\p. \eqno{(1.4)}$$

{\em Proof.} From (1.2) we have $\hat w_{i+1}\p\hat w_{i+1}^{-1}(\p+v_i)=
(\p+v_i)\hat w_{i}\p\hat w_{i}^{-1}$ or $$\p\hw_{i+1}^{-1}(\p+v_i)\hw_{i}=
\hw_{i+1}^{-1}(\p+v_i)\hw_{i}\p,$$ i.e., $\hw_{i+1}^{-1}(\p+v_i)\hw_{i}$ 
commutes with $\p$ being a first order $\Psi$DO. It has a form
$\p(1+\sum_0^\infty c_{i,k}\p^{-k})$. If we replace $\hw_{i+1}$ by $\hw_{i}
(1+\sum_0^\infty c_{i,k}\p^{-k})$, then $\hw_{i+1}^{-1}(\p+v_i)\hw_{i}=\p$,
i.e., (1.4). Now we can start with some $i$ and improve in succession
$\hw_{i+1},\hw_{i+2},...,\hw_{i+n}$. We have $$\hw_{i+n}=(\p+v_{i+n-1})...
(\p+v_i)\hw_i\p^{-n}=L_iL_i^{-1}\hw_i=\hw_i;$$ and $\hw_i$ depends on the 
index $i$ periodically, like $L_i$. $\Box$

It is not difficult to show that $\p_i$ defined by (1.3) commute.

Baker functions $w_i(t,z)$ corresponding to the dressing operators
$\hw_i(t,\p)$ are $w_i(t,z)=\hw_i(t,\p)\exp\xi(t,z)$ where $\xi(t,z)=
\sum_1^\infty t_iz^i$. The Eq. (1.4) is equivalent to $$(\p+v_i)w_{i}(t,z)
=zw_{i+1}(t,z). \eqno{(1.5)}$$

{\bf Remark.} It is worth mentioning that in terms of Grassmannians the
relations (1.4) or (1.5) mean the following. If $V_i$ are elements of
the Grassmannian related to $w_i$ then $zV_{i+1}\subset V_{i}$ (see [D 93]).
Grassmannians considerations help to build examples; for instance, let
$H$ be the space $L_2$ on the circle $|z|=1$, and 
$$V_i=\{f(z)=\sum_{-N}^{\infty}f_kz^k\:|\:f(a_l)=\epsilon^{i}\alpha_lf(
\epsilon a_l);\:l=1,...,N,\:\epsilon^n=1\}. \eqno{(1.6)}$$ Functions $f$ are 
supposed to be prolonged into the circle and $a_l$ are distinct points, 
$0<|a_l|<1$, while $\a_l$ are arbitrary non-zero numbers. It is easy to see 
that all properties are satisfied. \\

A transition from one solution $L_i$ of KdV to the others, $L_j$, is a 
B\"acklund (or Darboux) transformation, see Adler [A 81].\\

{\bf 2. Modified KP.}\\

There exists several definitions of this hierarchy given by various authors. 
The first was suggested by Kuperschmidt [K 89], see also
Yi Cheng [Y 93] and Gestezy and Unterkofler [GU 95]. All the definitions are 
trying to transfer 
the relationship between KdV and mKdV to the KP situation and, first of all,
to factorize the KP operator. And there is a big obstacle on this way, even 
insurmountable one. A pseudodifferential operator cannot be represented as
a product of first order differential operators. Suggested palliatives like
a product of a finite number of operators with one $\Psi$DO factor and the
rest of them being first order differential operators have a disadvantage
being not symmetric and not allowing a B\"acklund transformation.. 
We abandon the very idea of the factorization and we will base our definition on
a concept of a collection of KP operators connected by the equation (1.2) (see
below). There are too many variables here, all of them cannot be independent.
The first problem is to find a complete set of independent variables. We suggest
the following construction.

Let $L_0$ be a KP operator and $v_i$ where $i\in {\bf Z}$ variables. For $i>0$
let $$L_i=(\p+v_{i-1})...(\p+v_0)L_0(\p+v_0)^{-1}...(\p+v_{i-1})^{-1}$$ and
$$L_{-i}=(\p+v_{-i})^{-1}...(\p+v_{-1})^{-1}L_0(\p+v_{-1})...(\p+v_{-i}).$$
Evidently, $$L_{i+1}(\p+v_i)=(\p+v_i)L_{i},~i\in{\bf Z}.\eqno{(2.1)}$$

Thus, we take the collection of all coefficients of $L_0$ along with the set 
of all $v_i$ as independent variables. Instead, we could take
coefficients of some other $L_{i_0}$ and the same set of $v_i$; it would
be a different system of variables.\\

{\bf Definition 2.1.} {\sl The modified KP hierarchy is a system of equations}:
$$\p_kL_0=[(L_0^k)_+,L_0],~~\p_kv_i=(L_{i+1}^k)_+(\p+v_i)-(\p+v_i)(L_{i}^{k})_+.
\eqno{(2.2)}$$ 

{\bf Proposition 2.2.} {\sl For all $i\in{\bf Z}$ the equation
$$ \p_kL_i=[(L_i^k)_+,L_i] \eqno{(2.3)}$$ holds.}\\

{\em Proof.} We use induction. Let $i>0$ and for all smaller non-negative
indices the eq. (2.3) be proven. We have: $$L_i=(\p+v_{i-1})L_{i-1}(\p+v_
{i-1})^{-1}$$ and $$\p_kL_i=-(\p+v_{i-1})L_{i-1}(\p+v_{i-1})^{-1}\{(L_{i}^k)_+(\p+v_{i-1})-
(\p+v_{i-1})(L_{i-1}^k)_+\}(\p+v_{i-1})^{-1}$$
$$+\{(L_{i}^k)_+(\p+v_{i-1})-(\p+v_{i-1})(L_{i-1}^k)_+\}L_{i-1}(\p+v_{i-1})
^{-1}$$ $$+(\p+v_{i-1})\left[(L_{i-1}^k)_+,L_{i-1}\right](\p+v_{i-1})^{-1}$$ 
$$=-L_{i}(L_{i}^k)_++(\p+v_{i-1})L_{i-1}(L_{i-1}^k)_+(\p+v_{i-1})^{-1}$$ 
$$+(L_{i}^k)_+L_{i}-(\p+v_{i-1})(L_{i-1}^k)_+L_{i-1}(\p+v_{i-1})^{-1}
$$ $$+(\p+v_{i-1})\left[(L_{i-1}^k)_+,L_{i-1}\right](\p+v_{i-1})^{-1}=
[(L_i^k)_+,L_i].$$ Similarly, $$L_{-i}=(\p+v_{-i})^{-1}L_{-i+1}(\p+v_{-i})$$
and $$\p_kv_{-i}=(L_{-i+1}^k)_+(\p+v_{-i})-(\p+v_{-i})(L_{-i}^k)_+.$$
Then, 
$$\p_kL_{-i}=-(\p+v_{-i})^{-1}\{(L_{-i+1}^k)_+(\p+v_{-i})-(\p+v_{-i})(L_{-i}^k)_+\}
(\p+v_{-i})^{-1}L_{-i+1}(\p+v_{-i})$$ $$+(\p+v_{-i})^{-1}L_{-i+1}
\{(L_{-i+1}^k)_+(\p+v_{-i})-(\p+v_{-i})(L_{-i}^k)_+\}
$$ $$+(\p+v_{-i})^{-1}[L_{-i+1}^k)_+,L_{-i+1}](\p+v_{-i})$$
$$=-(\p+v_{-i})^{-1}(L_{-i+1}^k)_+L_{-i+1}(\p+v_{-i})+(L_{-i}^k)_+L_{-
i}$$ $$+(\p+v_{-i})^{-1}L_{-i+1}(L_{-i+1}^k)_+(\p+v_{-i})-L_{-i}(L_{-
i}^k)_+$$ $$+(\p+v_{-i})^{-1}[(L_{-i+1}^k)_+,L_{-i+1}](\p+v_{-i})=[(L_{-i}^k)_+,
L_{-i}].~\Box$$

{\bf Proposition 2.3.} {\sl Derivations $\p_k$ commute.}\\

{\em Proof.} The fact that $\p_k\p_lL_i=\p_l\p_kL_i$ is known; this is a property
of the KP hierarchy. Now, $$\p_k\p_lv_i-\p_l\p_kv_i=\p_k
\left\{(L_{i+1}^l)_+(\p+v_i)-(\p+v_i)(L_{i}^l)_+\right\}-(k\leftrightarrow l)$$
$$=[(L_{i+1}^k)_+,L_{i+1}^l]_+(\p+v_i)-(\p+v_i)[(L_{i}^k)_+,L_{i}^l]_+$$
$$+(L_{i+1}^l)_+\left\{(L_{i+1}^k)_+(\p+v_i)-(\p+v_i)(L_{i}^k)_+\right\}$$ $$
-\left\{(L_{i+1}^k)_+(\p+v_i)-(\p+v_i)(L_{i}^k)_+\right\}(L_{i}^l)_+-
(k\leftrightarrow l)$$ $$=\left\{(L_{i+1}^k)_+L_{i+1}^l-(L_i^l)_-L_i^k\right\}_+
(\p+v_i)-(\p+v_i)\left\{L_{i}^k(L_{i}^l)_--L_{i}^l(L_{i}^k)_+\right\}
-(k\leftrightarrow l)$$ $$=\left\{L_{i+1}^kL_{i+1}^l-L_{i+1}^lL_{i+1}^k\right\}_+(\p+v_i)
-(\p+v_i)\left\{L_{i}^kL_{i}^l-L_{i}^lL_{i}^k\right\}=0.~\Box$$

Dressing operators can be introduced: $$\hw_i(t,\p)=\sum_{-\infty}^0w_{i\a}\p^\a,
~w_{i0}=1 \eqno{(2.4)}$$ such that 
$$ L_i=\hw_i\p\hw_i^{-1},~~(\p+v_i)\cdot \hw_{i}=\hw_{i+1}\cdot\p, \eqno{(2.5)}$$
Baker functions $$w_i(t,z)=\hw_i(t,\p)\exp\xi(t,z),L_iw_i(t,z)=zw_i(t,z),
~(\p+v_i)w_{i}(t,z)=zw_{i+1}(t,z) \eqno{(2.6)}$$ and conjugate Baker functions:
$$w^*_i(t,z)=(\hw_i(t,\p)^{-1})^*\exp(-\xi(t,z)),~L^*_iw^*_i(t,z)=zw^*_i(t,z),~
(\p-v_i)w_{i+1}^*(t,z)=-zw^*_{i}(t,z). \eqno{(2.7)}$$ Here the asterisk * in $w_i^*$
just belongs to the notation while in $\hw_i^*$ and $L_i^*$ it means
formal conjugate of operators.\\

{\bf Proposition 2.4.} {\sl There is an authomorphism of the mKP:}
$$t_k\mp \tilde t_k=(-1)^{k-1}t_k,~v_i\mp \tilde v_i=-v_{-i-1},~ L_i\mp \tilde L_i=
 -L_{-i}^*,$$ $$\hw_i\mp\tilde{\hw}_i=(\hw_{-i}^{-1})^*,~w_i\mp \tilde
w_i=w_{-i}^*,~ z\mp\tilde z=-z.\eqno{(2.8)}$$

Indeed, it is easy to see that the equations defining the hierarchy
tolerate this transformation. $\Box$

Notice that any streak of equations (2.2), finite of semi-infinite,
$0\leq i<i_1$ or $i_1<i\leq 0$ (in particular, one equation, $i=0$) form a closed 
system. Especially interesting are the semi-infinite cases, $0\leq i<\infty$ or 
$-\infty<i\leq 0$. These are one-sided mKP's: mKP$_+$ and mKP$_-$. The 
authomorphism (2.8) interchanges them. 

Also notice that the mKdV is a restriction of mKP to the case when
$v_{i+n}=v_i$ and $L_0=(\p+v_{n-1})...(\p+v_0)$.\\

{\em The rest of this section is not used in the proof of the main theorem
(3.7) and can be skipped.}\\

\begin{small}
It is well-known that the so-called Hirota-Sato bilinear identity gives an 
equivalent description of the KP hierarchy. It is based on the following
fundamental lemma (see [D 91] or [D 97])):\\

{\bf Lemma 2.5.} {\sl Let $P$ and $Q$ be two $\Psi$DO,
then $${\rm res}_z~[(Pe^{xz})\cdot(Qe^{-xz})]={\rm res}_{\p}~PQ^*$$ where $Q^*$
is the formal adjoint to $Q$.}\\

{\em Proof}. The left-hand side is $${\rm res}_z~[(Pe^{xz})\cdot(Qe^{-xz})]=
{\rm res}_z~(\sum p_iz^i\sum q_j(-z)^j)=\sum_{i+j=-1}(-1)^jp_iq_j,$$ and the
right-hand side is $${\rm res}_{\p}~(P\cdot Q^*)=\res_{\p}~\sum_{ij}p_i\p^i(-\p)
^jq_j=\sum_{i+j=-1}p_iq_j.~\Box$$

{\bf Proposition 2.6 (bilinear identity).} {\sl If $w$ is a Baker function of
the KP hierarchy, then the identity $${\rm res}_z~(\p_1^
{k_1}...\p_m^{k_m}w)\cdot w^*=0$$ holds for any $(k_1,...,k_m)$.}\\

{\em Proof.} Since $\p_kw=L_+^kw$, it suffices to consider only the case when 
$m=1$. Then $${\rm res}_z~(\p^kw)\cdot w^*={\rm res}_z~(\p
^k\hw e^{\xi(t,z)})(\hw^*)^{-1}e^{-\xi(t,z)}$$ $$={\rm res}_z~(\p^k\hw e^{xz})((
\hw^*)^{-1}e^{-xz})={\rm res}_{\p}~\p^k\hw\cdot\hw^{-1}={\rm res}_{\p}~\p^k=0.
~\Box$$
Notice that if the hierarchy equations are not used and it is only assumed
that $w$ and $w^*$ are connected by the formulas $w=\hw\exp\xi$ and $w^*=(\hw^*)^
{-1}\exp(-\xi)$ then a weak form of bilinear identity will be obtained, res$_z\p
^kw\cdot w^*=0$. 

Return to our mKP hierarchy.\\

{\bf Proposition 2.7.} {\sl If $w_i$ are Baker functions of the mKP hierarchy
then the identity $${\rm res}_z~(z^{i-j}\p_1^
{k_1}...\p_m^{k_m}w_i)\cdot w_j^*=0, ~{\rm when}~i\geq j\eqno{(2.9)}$$ holds for any 
$(k_1,...,k_m)$.}\\

{\em Proof.} Again, it suffices to take $m=0$. Then
$${\rm res}_z~z^{i-j}(\p^kw_i)\cdot w_j^*={\rm res}_z~(\p
^k\hw_i\p^{i-j} e^{\xi(t,z)})(\hw_j^*)^{-1}e^{-\xi(t,z)}$$ $$={\rm res}_z~(\p^k
(\p+v_{i-1})...(\p+v_j)\hw_j e^{xz})((\hw_j^*)^{-1}e^{-xz})$$ 
$$={\rm res}_{\p}~\p^k(\p+v_{i-1})...(\p+v_j)\hw_j\cdot\hw_j^{-1}=
{\rm res}_{\p}~\p^k(\p+v_{i-1})...(\p+v_j)=0.~\Box$$

The converse is also true.\\

{\bf Proposition 2.8.} {Let $$w_i=\sum_0^\infty w_{i\a}z^{-\a}e^{\xi(t,z)},~w_i^*=
\sum_0^\infty w_{i\a}^*z^{-\a}e^{-\xi(t,z)}$$ be formal expansions where $w_{i\a}$ 
and $w_{i\a}^*$ are functions of variables $t_k$, and $w_{i0}=w_{i0}^*=1$. Let 
$${\rm res}_z~z^{i-j}(\p_1^{k_1}...\p_m^{k_m}w_i)\cdot w_j^*=0,~{\rm when}~
i=j,j+1$$ hold for any multiindex. Then 
$w_i$ and $w_i^*$ are the Baker and the adjoint Baker functions of 
the mKP hierarchy.}\\

{\em Proof.} When $i=j$, this is a well-known theorem that states that
all $w_i$s and $w_i^*$s are Baker and conjugate Baker functions of KP (see, 
e.g., [D 91] or [D 97]). When $i=j+1$ we have
$$0=\res_z~z\p^kw_{j+1}w_j^* =\res_z~\p^k\hw_{j+1}\p e^\xi (\hw_j^*)^{-
1}e^{-\xi}=\res_\p~\p^k\hw_{j+1}\p\hw_j^{-1}$$ which yields that
$(\hw_{j+1}\p\hw_j^{-1})_-=0$ and $\hw_{j+1}\p\hw_j^{-1}$ is a 
first-order differential operator: $\hw_{j+1}\p\hw_j^{-1}=\p+v_i$. Then
$$(\p+v_i)\hw_j=\hw_{j+1}\p,$$ and this proves the proposition. $\Box$\\

\end{small}
{\bf 3. Discrete KP.} \\

We deal with a linear space $H$ of $\Psi$DO: \{$a=\sum_{-\infty} a_i\p^i$\}; the
series are one-way infinite. A dual space $H^*$ is \{$b=\sum_{-\infty}
\p^{-i-1}b_i$\}; a coupling is given by $<a,b>=\res_\p~ab=\sum_{-\infty}a_ib_i$.

Infinite matrices $\bM=(m_{ij})$ where $i,j\in{\bf Z}$ will be considered as matrices
of a change of the basis: $\bM_i=\sum_{\a=-\infty}m_{i\a}\p^{\a}$ are new basis vectors,
instead of $\p^i$. The same matrices also can be treated as matrices of
basis change in the dual space, $\bM^j=\sum ^{\b=\infty}\p^{-\b-1}m_{\b j}$ being new
basis vectors instead of $\p^{-j-1}$. Product $\bP=\bM\bN$ of two matrices can
be found as $\bP_{ij}=\res_\p \bM_i\bN^j$. All the matrices will be triangular,
$m_{ij}=0$ when $i>j+$const, and all the sums make sense. Thus, the rows are
associated with operators $\bM_i$ and the columns with operators $\bM^j$\footnote{In the
same manner we treated the Drinfeld-Sokolov reduction in [D 81], see also [D
91] or [D 97]: differential operators $\bM_i=\sum_\a m_{i\a}\p^{\a}$ were
associated with the rows and integral operators  $\bM^j=\sum _\b\p^{-\b-1}m_{\b j}$
with the columns of finite matrices.}.

Let $\bW$ be a matrix with rows $\bW_i=\hw_i\p^i$ where $\hw_i$ are dressing operators
(2.4). Let $$\hw_j^{-1}=\sum_{\b=0}^\infty\p^{-\b}\tilde w_{\b j},~
w_{0j}=1\eqno{(3.1)}$$ and let $\tW$ be a matrix with columns $\tW^j=\p^{-j-
1}\hw_{j+1}^{-1}$.\\

{\bf Lemma 3.1.} $\tW=\bW^{-1}$.\\

 {\em Proof.} The matrix elements of $\bW$ are $\bW_{ij}=w_{i,j-i}$
when $i\geq j$ and 0 otherwise. The matrix elements of $\tW$ are
$\tW_{ij}=\tilde w_{i-j,j+1}$ if $i\geq j$ and 0 otherwise. Both, $\bW$ and
$\tW$, are lower triangular matrices with unities on their diagonals. So is 
their product $\bW\tW$, its subdiagonal elements are $$(\bW\tW)_{ij}=\res_\p \bW_i\tW^j=
\res_\p\hw_i\p^i\p^{-j-1}\hw_{j+1}^{-
1},~~i>j.$$ Eq. (2.5) implies that if $i>j$ then $$\hw_i=(\p+v_{i-1})\hw_{i-1}
\p^{-1}=...=(\p+v_{i-1})...(\p+v_{j+1})\hw_{j+1}\p^{-i+j+1}$$ whence
$$ (\bW\tW)_{ij}=\res_\p(\p+v_{i-1})...(\p+v_{j+1})\hw_{j+1}\p^{-i+j+1}\p^{i-j-
1}\hw_{j+1}^{-1}=\res_\p(\p+v_{i-1})...(\p+v_{j+1})=0.$$ Thus, $\bW\tW=I$, the
matrix unity, and $\tW=\bW^{-1}$. $\Box$

{\bf Definition 3.2.} {\sl If $\bM$ is a matrix then $\bM_+$ is a matrix
such that $$(\bM_+)_{ij}=\left\{\begin{array}{rr}\bM_{ij},&{\rm when}~i\geq
j\\0,&{\rm when}~i<j\end{array}\right.$$ and $\bM_-=\bM-\bM_+$.}\\

It is easy to see that $$(\bM_+)_{i}=(\bM_i\p^{-i})_+\p^i,~~(\bM_-)_{i}=
(\bM_i\p^{-i})_-\p^i.\eqno{(3.2)}$$

Let $\L$ be the matrix $\L_{ij}=\d_{i,j-1}$.\\

{\bf Lemma 3.3.} {\sl If $\bM=(M_{ij})$ is a matrix then an operator associated
with a row of a matrix $\bM\L$ is $\bM_i\p$.}\\

{\em Proof.} $$(\bM\L)_i=\sum_j (\bM\L)_{ij}\p^j=\sum_j\bM_{i,j-1}
\p^j=\sum_j\bM_{i,j-1}\p^{j-1}\p=\bM_i\p.~\Box$$

Let us dress the matrix $\L$ with the help of $\bW$: $\bL=\bW\L \bW^{-1}$.\\

{\bf Proposition 3.4.} {\sl By virtue of the mKP equations (2.2),
the following is true}: $$\p_k\bW=-(\bL^k)_-\bW. \eqno{(3.3)}$$

{\em Proof.} $$\p_k\bW_i=\p_k\hw_i\p^i=-(L_i^k)_-\hw_i\p^i,$$ $$(\p_k\bW_i\cdot
\bW^{-1})_{ij}=\res_\p(-(L_i^k)_-\hw_i\p^i\p^{-j-
1}\hw_{j+1}^{-1})=-\res_\p(L_i^k)_-(\p+v_{i-1})...(\p+v_{j+1}).$$
On the other hand, $$(\bL^k)_{ij}=(\bW\L^k\bW^{-1})_{ij}=\res_\p(\bW\L^k)_i(\bW^{-
1})^j=\res_\p\hw_i\p^i\p^k\p^{-j-1}\hw_{j+1}^{-1}$$ $$=\res_\p L_i^k\hw_i
\p^{i-j-1}\hw_{j+1}^{-1}=\res_\p L_i^k(\p+v_{i-1})...(\p+v_{j+1})=
\res_\p(L_i^k)_-(\p+v_{i-1})...(\p+v_{j+1}).~\Box$$

{\bf Corollary 3.5.} {\sl The mKP equations (2.2) imply}
$$\p_k\bL=[(\bL^k)_+,\bL]. \eqno{(3.4)}$$

{\em Proof.} $$\p_k\bL=\p_k(\bW\L \bW^{-1})=-(\bL^k)_-\bW\L \bW^{-1}+\bW\L \bW^{-
1}(\bL^k)_-\bW\bW^{-1}$$ $$=-[(\bL^k)_-,\bL^k]=[(\bL^k)_+,\bL]. ~\Box$$

{\bf Definition 3.6.} {\sl The equation (3.4) where $\bL=\L+$(lower
triangular) is the discrete KP.}\\

{\bf Theorem 3.7.} {\sl The discrete KP (3.4) is equivalent to the
modified KP (2.2).}\\

{\em Proof.} Since this is already proven in one direction, it remains to show 
that each solution to (3.4) can be obtained from a solution to (2.2) in the
above described way.

Thus, given a matrix $\bL=\L+$(lower triangular), satisfying (3.4). The
matrix $\bL$ can represented in a form $\bL=\bW\L\bW^{-1}$. 
Lower triangular dressing matrices $\bW$ with unities on the main diagonal are 
not uniquely determined; they can be multiplied on the right by a 
matrix commuting with $\L$, i.e., constant on all diagonals. Using this freedom,
it is always possible to satisfy (3.3) (there remains a little freedom even 
after this operation: $\bW$ can be multiplied on the right by a constant matrix
commuting with $\bL$, this can be fixed with the initial conditions). Thus, one
can consider (3.3) as a discrete KP equation. The variable $t_1$ we will
identify with $x$ and $\p_1=\p$.

Let $\bW_i=\hw_i\p^i$ be operators
associated with rows of $W$ and $(\bW^{-1})^j=\p^{-j-1}\tilde w_{j+1}$ operators
associated with columns of $\bW^{-1}$. In the next three lemmas there will be assumed
that $\bW$ satisfies (3.3).  \\

{\bf Lemma 3.8.} {\sl There exist quantities $v_i$ such that} $$
(\p+v_i)\hw_i=\hw_{i+1}\p.$$

{\em Proof.} The equation (3.3) for $k=1$ reads $$\p_1W=-\bL\bW+\bL_+\bW=
-\bW\L+(\L-V)\bW$$ where $V$ is a diagonal matrix. Let us take operators
associated with the $i$th row on the left and on the right of this equality
taking into account the Lemma 3.3. $\p(\hw_i\p^i)=-\hw_i\p\p^i
+\hw_{i+1}\p^{i+1}-v_i\hw_i\p^i$ or $(\p+v_i)\hw_i=\hw_{i+1}\p$. $\Box$\\

{\bf Lemma 3.9.} $$\tilde w_{j+1}=w_{j+1}^{-1}.$$          
											
{\em Proof.} We have $(\bW\bW^{-1})_{ij}=0$ when $i>j$, therefore 
$$0=\res_\p~\hw_i\p^i\p^{-j-1}\tilde w_{j+1}=\res_\p~(\p+v_{i-1})...(\p+v_{j+1})
\hw_{j+1}\tilde w_{j+1}.$$ This can be true for all $i>j$ only if $
(\hw_{j+1}\tilde w_{j+1})_-=0$. Then $\hw_{j+1}\tilde w_{j+1}=1$ and
$\tilde w_{j+1}=\hw_{j+1}^{-1}$. $\Box$\\

{\bf Lemma 3.10.} {\sl The operators $\hw_i$ satisfy KP.}\\

{\em Proof.} The $ij$th element of the matrix equality $\p_k\bW\cdot\bW^{-1}
=-(\bW\L^k\bW^{-1})_-$ is
$$\res_\p~\p_k(\hw_i)\p^i\p^{-j-1}\hw_{j+1}^{-1}=-\res_\p~\hw_i\p^k\p^{i-j-1}
\hw_{j+1}^{-1},~i>j$$or $$\res_\p~(\p_k\hw_i+\hw_i\p^k\hw_i^{-1})
(\p+v_{i-1})...(\p+v_{j+1})=0$$ fir a given $i$ and all $j<i$. This implies
$(\p_k\hw_i+\hw_i\p^k\hw_i^{-1})_-=0$ and, finally, $
\res_\p~\p_k\hw_i\cdot\hw_i^{-1}=-(\hw_i\p^k\hw_i^{-1})_-$ which is KP. $\Box$\\

This also completes the proof of the theorem since $(\p+v_i)=\hw_{i+1}\p\hw_i$
and $$\p_kv_i=-(L_{i+1}^k)_-(\p+v_i)+(\p+v_i)(L_i^k)_-=(L_{i+1}^k)_+(\p+v_i)-
(\p+v_i)(L_i^k)_+.~\Box$$ 

{\bf References.}\\

[A 81] Adler, M., On the B\"acklund transformation for the
Gelfand-Dikii equations, Comm. Mat. Phys., 80, no 4, 517-527, 1981\\

[AvM 98] Adler, M. and van Moerbecke, P., Vertex operator solutions to the
discrete KP-hierarchy, Preprint, 1998\\

[D 81] Dickey, L.A., Integrable nonlinear equations and Liouville's 
theorem I, Comm. Math. \mbox{} Phys., 82, no3, 345-360, 1981\\

[D 91] Dickey, L.A., Soliton equations and Hamiltonian systems, Advanced Series 
in Mathematical Physics, Vol. 12, World Scientific, 1-310, 1991\\

[D 93] Dickey, L.A., Additional symmetries of KP, Grassmannian, and the string 
equation II, Modern Physics Letters A, 8, no14, 1357-1377, 1993\\

[D 94] Dickey, L.A., Why the general Zakharov-Shabat equations form a hierarchy, 
Com. Math. Phys., 163, no3, 509-522, 1994\\

[D 97] Dickey, L.A., Lectures on classical W-algebras, Acta Applicandae 
Mathematicae, 47, 243-321, 1997\\

[DS 81] Drinfeld, V.G. and Sokolov V.V., Equations of
the Korteweg-de-Fries type and simple Lie algebras, Doklady AN SSSR,
258, no1, 11-16, 1981 \\

[GU 95] Gestezy, F. and Unterkofler, K., On the (modified) Kadomtsev-Petviashvili 
hierarchy, Differential and Integral Equations, 8, no. 4,
797-812,1995\\

[K 89] Kupershmidt, B.A., On the integrability of modified Lax euations,
Journ. Phys. A, 22, L993, 1989\\

[UT 84] Ueno, K. and Takasaki, K., Toda lattice hierarchy, Adv. Studies
in Pure Math. 4, 1-95, 1984\\

[Y 93] Yi Cheng, Modifying the KP, the Nth constrained KP hierarchies
and their Hamiltonian structures, preprint, 1993\\

\end{document}